\documentclass[12pt]{iopart}

\usepackage{bm}
\usepackage{graphicx}
\begin{document}
\title[R G Zhu]{The single-ion anisotropy in LaFeAsO}

\author{Ren-Gui Zhu}

\address{College of physics and
electronic information, Anhui Normal University, Wuhu, 241000, P. R.
China} \ead{rgzhu@mail.ahnu.edu.cn}
\begin{abstract}
We use Green's function method to study the Heisenberg model of
LaFeAsO with the striped antiferromagnetic collinear spin structure.
In addition to the intra-layer spin couplings $J_{1a},J_{1b},J_2$
and the inter-layer coupling $J_c$, we further consider the
contributions of the single-ion anisotropy $J_s$. The analytical
expressions for the magnetic phase-transition temperature $T_N$ and
the spin spectrum gap $\Delta$ are obtained. According to the
experimental temperature $T_N=138$K and the previous estimations of
the coupling interactions, we make a further discussion about the
magnitude and the effects of the single-ion anisotropy $J_s$. We
find that the magnitudes of $J_s$ and $J_c$ can compete. The
dependences of the transition temperature $T_N$, the
zero-temperature average spin and the spin spectrum gap on the
single-ion anisotropy are investigated. We find they both increase
as $J_s$ increases. The spin spectrum gap at low temperature $T\to0$
is calculated as a function of $J_s$, the result of which is a
useful reference for the future experimental researches.
\end{abstract}

\maketitle

\section{Introduction}
It was recently discovered that an iron-based material LaFeAsO shows
high-temperature superconductivity when O atoms are partially
substituted by F atoms\cite{Kami}. This discovery has triggered
great research interest on the FeAs-based pnictides superconductors
and their undoped compounds. It has been theoretically and
experimentally confirmed that these pure FeAs-based compounds have a
ground state with collinear stripe-like antiferromagnetic(AF) spin
order formed by Fe
atoms\cite{Yild,Cao,Ma,Dong,Cruz,Chen,Huang,Zhao2,Gold,McGu}. Thus
to establish an effective spin Hamiltonian for them and to elucidate
the corresponding antiferromagnetism are helpful in understanding
the underlying mechanism to make them superconducting upon doping.

\begin{figure}
  \begin{center}
  \includegraphics[scale=0.6]{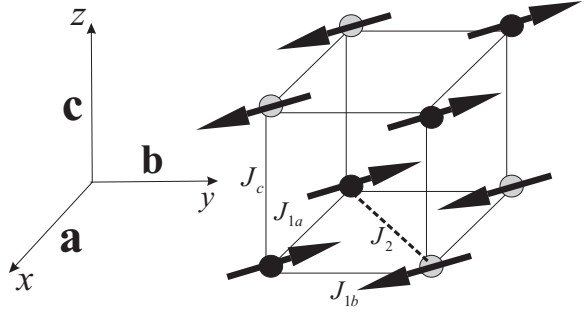}
  \end{center}
  \caption{\label{fig1}A unit cell of the orthorhombic Fe spin lattice. $\bm a$, $\bm b$,
  $\bm c$ are the three base vectors. The lattice consists of two sublattices
  (distinguished by the color gray and black).}
\end{figure}

 For undoped LaFeAsO and other similar parent compounds, a Heisenberg exchange model was suggested to
 explain their AF structure\cite{Yild,Ma,Si,Fang}, and was used to
 explore their magnetic properties\cite{Zhao,Liu,Yao}. Figure \ref{fig1} shows the unit cell of the orthorhombic
 AF spin structure of the Fe lattice. This orthorhombic structure
 exists below a structure transition temperature $T_S$, which is
 $15\sim 20$K higher than the magnetic transition temperature $T_N$\cite{Cruz,Klau}.
 Usually the nearest neighbor (NN) coupling $J_1$(including $J_{1a}$ and $J_{1b}$),
 and the next-nearest neighbor (NNN) coupling $J_2$ in FeAs
 layers are dominant and must be considered. The NN coupling $J_c$ between
 spins on neighboring layers is regarded to be much smaller than the
 in-plane couplings\cite{Yild,Ma}. However, $J_c$ was found to be
 essential for the existence of a non-zero magnetic transition
 temperature $T_N$\cite{Liu}. A further consideration
 can include the single-ion anisotropy $J_s$. It was estimated to be even much smaller
than $J_c$ in the model of SrFe$_2$As$_2$, but a spin spectrum gap
was found to be produced by it\cite{Zhao}. For LaFeAsO, so far there
is no research report about the magnitude or the effects of the
single-ion anisotropy.

In this paper, we use Green's function method\cite{Tyab} to study
the Heisenberg model of FeAs-based pure parent compounds. The
Hamiltonian of this model in a detailed form is
\begin{eqnarray}
H&=&\frac{1}{2}J_{1a}\sum_{\langle ij\rangle}\bm S_{1i}\cdot\bm
S_{1j}+\frac{1}{2}J_{1a}\sum_{\langle ij\rangle}\bm S_{2i}\cdot\bm
S_{2j}+J_{1b}\sum_{\langle ij\rangle}\bm S_{1i}\cdot\bm
S_{2j}\nonumber\\
&&+J_2\sum_{\langle\langle ij\rangle\rangle}\bm S_{1i}\cdot\bm
S_{2j}+J_c\sum_{\langle ij\rangle}\bm S_{1i}\cdot\bm
S_{2j}-J_s\sum_i[(S_{1i}^z)^2+(S_{2i}^z)^2],\label{eq01}
\end{eqnarray}
where the spin coupling $J_c$ between layers and the single-ion
anisotropy $J_s$ are both considered. The subscripts 1 and 2 mean
the sublattices 1 and 2 respectively. $\langle ij\rangle$ means NN
spin pairs, and $\langle\langle ij\rangle\rangle$ means NNN spin
pairs. The self-consistent equations for the average sublattice spin
will be derived. An analytical expression for the magnetic
transition temperature $T_N$ will be obtained. For LaFeAsO,
according to the recent estimations of the strengths $J_{1a},J_{1b},
J_2$ and $J_c$\cite{Liu} with the experimental temperature
$T_N=138$K\cite{Cruz,Klau},  we shall make a further estimation of
the single-ion anisotropy $J_s$. We find that the magnitude of $J_s$
can compete with $J_c$, and in some situations even bigger than
$J_c$. The effects of the single-ion anisotropy on the transition
temperature $T_N$, the zero-temperature average spin $\langle
S_z\rangle_0$ and the spin spectrum gap are investigated. We find
they all increase as $J_s$ increases. In section \ref{sec2}, we
shall give our analytical results derived from the Green's function
method. In section \ref{sec3}, we shall present our numerical
results. Finally a conlusion is given in section \ref{sec4}.
\section{Green's function derivation}\label{sec2}
According to the general scheme of Green's function method to solve
an antiferromagnetic spin model with two sublattices, we construct
the following Green's functions:
\begin{equation}
G_{1k,1l}(\omega)=\langle\langle S_{1k}^+;S_{1l}^-\rangle\rangle,\
G_{2k,1l}(\omega)=\langle\langle S_{2k}^+;S_{1l}^-\rangle\rangle.
\end{equation}
The equation of motion is
\begin{equation}
\omega\langle\langle
A;S_{1l}^-\rangle\rangle=\langle[A,S_{1l}^-]\rangle+\langle\langle
[A,H];S_{1l}^-\rangle\rangle,
\end{equation}
where $A$ represents the spin operator $S_{1k}^+$ or $S_{2k}^+$. The
commutator $[A,H]$ can be derived using Hamiltonian (\ref{eq01}) and
the basic commutation relations of spin
operators:$[S_i^+,S_j^-]=2S_i^z\delta_{ij}$, $[S_i^z,S_j^{\pm}]=\mp
S_i^{\pm}\delta_{ij}$, where $S_i^{\pm}=S_i^x\pm iS_i^y$.

In order to close the system of equations, the so-called PRA or
Tyablikov decoupling\cite{Tyab} is adopted for the terms stemming
from the exchange couplings:
\begin{equation}
\langle\langle S_i^zS_j^+;S_l^-\rangle\rangle\approx\langle
S_i^z\rangle\langle\langle S_j^+;S_l^-\rangle\rangle,\ \ i\ne j.
\end{equation}
While for the terms stemming from the single-ion anisotropy, we
adopt the Anderson-Callen(AC) decoupling\cite{Ander}:
\begin{equation}
\langle\langle S_i^zS_i^++
S_i^+S_i^z;S_j^-\rangle\rangle\approx2\langle
S_i^z\rangle\Theta_i^{(z)}\langle\langle S_i^+;S_j^-\rangle\rangle,
\end{equation}
where
\begin{equation}\label{eq04}
\Theta_i^{(z)}=1-\frac{1}{2S^2}[S(S+1)-\langle S_i^zS_i^z\rangle].
\end{equation}
The AC decoupling has been demonstrated to be most adequate for the
single-ion anisotropy much small compared to the exchange
interactions\cite{Frob,Hene}.

In order to write the decoupled equations of motion in the $\bm k$
space, we take the following Fourier transformation:
\begin{equation}
G_{k,l}(\omega)=\frac{1}{N}\sum_{\bm k}G(\bm k,\omega)e^{i\bm
k\cdot(\bm R_k-\bm R_l)},
\end{equation}
where $N$ is the number of sites in either sublattice, and the
summation over $\bm k$ is restricted to the first Brillouin zone of
the sublattice. At the same time, the equation
$\delta_{ij}=\frac{1}{N}\sum_{\bm k}e^{i\bm k\cdot(\bm R_i-\bm
R_j)}$ is also used.

Because of the translation invariant, we have $\langle
S_{1k}^z\rangle=\langle S^z\rangle,\ \langle
S_{2k}^z\rangle=-\langle S^z\rangle$ and
$\Theta_{1k}^{(z)}=\Theta_{2k}^{(z)}=\Theta^{(z)}=1-\frac{1}{2S^2}[S(S+1)-\langle
S^zS^z\rangle]$. Finally, we obtain the decoupled equations of the
two Green's function in $\bm k$ space:
\begin{equation}\label{eq02}
[\omega-\langle S^z\rangle A_{\bm k}]G_{11}(\bm k,\omega)-\langle
S^z\rangle B_{\bm k}G_{21}(\bm k,\omega)=2\langle S^z\rangle,
\end{equation}
and
\begin{equation}\label{eq03}
[\omega+\langle S^z\rangle A_{\bm k}]G_{21}(\bm k,\omega)+\langle
S^z\rangle B_{\bm k}G_{11}(\bm k,\omega)=0,
\end{equation}
where
\begin{equation}
A_{\bm k}
=2J_{1a}\cos(k_xa)-2J_{1a}+2J_{1b}+4J_2+2J_c+2J_s\Theta^{(z)},
\end{equation}
and
\begin{equation}
B_{\bm k}=2J_{1b}\cos(k_yb)+4J_2\cos(k_xa)\cos(k_yb)+2J_c\cos(k_zc),
\end{equation}
in which $a,b,c$ are the three lattice constants. Solving equations
(\ref{eq02}) and (\ref{eq03}), we obtain the Green's function:
\begin{equation}
G_{11}(\bm k,\omega)=\frac{\langle S^z\rangle}{\omega_{\bm
k}}\left[\frac{A_{\bm k}\langle S^z\rangle+\omega_{\bm
k}}{\omega-\omega_{\bm k}}-\frac{A_{\bm k}\langle
S^z\rangle-\omega_{\bm k}}{\omega+\omega_{\bm k}}\right],
\end{equation}
and the spin spectrum:
\begin{equation}
\omega_{\bm k}=\langle S^z\rangle\sqrt{A_{\bm k}^2-B_{\bm k}^2}.
\end{equation}
When $\bm k\to0$, we obtain a expression for the spectrum gap:
\begin{equation}
\Delta=2\langle
S^z\rangle\sqrt{J_s\Theta^{(z)}[2J_{1b}+4J_2+2J_c+J_s\Theta^{(z)}]},
\end{equation}
which is similar with the expression given in ref\cite{Zhao} derived
from the spin-wave theory, expect for the factor $\Theta^{(z)}$.
From this expression for the gap, one see that the single-ion
anisotropy is essential for the existence of the spectrum gap.

Then following the process of solving the average spin, we derive
the correlation function $\langle S^-S^+\rangle$ using the spectrum
theorem:
\begin{eqnarray}
\langle S^-S^+\rangle&=&-\frac{1}{N\pi}\sum_{\bm
k}\int_{-\infty}^{\infty}d\omega\frac{\mbox{Im}G_{11}(\bm
k,\omega+i\epsilon)}{e^{\beta\omega}-1}\nonumber\\
&=&\frac{\langle S^z\rangle}{N}\sum_{\bm k}\left[\frac{A_{\bm
k}}{\sqrt{A_{\bm k}^2-B_{\bm k}^2}}\coth\frac{\beta\omega_{\bm
k}}{2}-1\right],
\end{eqnarray}
in which the equation
$\frac{1}{x+i\epsilon}=P(\frac{1}{x})-i\pi\delta(x)$($P(\cdots)$
means taking the principle value) has been used to obtain the
imaginary part of $G_{11}(\omega+i\epsilon)$, and
$\beta=\frac{1}{k_BT}$, $k_B$ is the Boltzmann constant, $T$ is the
temperature.

According to the theory of Callen\cite{Call}, the average spin for
arbitrary $S$ can be calculated using the following equation:
\begin{equation}\label{eq05}
\langle
S^z\rangle=\frac{(S-\Phi)(1+\Phi)^{2S+1}+(S+1+\Phi)\Phi^{2S+1}}{(1+\Phi)^{2S+1}-\Phi^{2S+1}},
\end{equation}
where
\begin{eqnarray}
\Phi&=&\frac{\langle S^-S^+\rangle}{2\langle
S^z\rangle}\nonumber\\
&=&\frac{1}{2N}\sum_{\bm k}\left[\frac{A_{\bm k}}{\sqrt{A_{\bm
k}^2-B_{\bm k}^2}}\coth\frac{\beta\omega_{\bm
k}}{2}-1\right].\label{eq06}
\end{eqnarray}
On the other hand, the correlation function$\langle S^zS^z\rangle$
can be calculated from the equation $\langle
S^zS^z\rangle=S(S+1)-(1+2\Phi)\langle S^z\rangle$. Using equation
(\ref{eq04}), we can relate $\Theta^{(z)}$ to $\Phi$ by
\begin{equation}\label{eq07}
\Theta^{(z)}=1-\frac{\langle S^z\rangle}{2S^2}(1+2\Phi).
\end{equation}
Now the equations (\ref{eq05})(\ref{eq06})(\ref{eq07}) can be solved
self-consistently to obtain the average spin at any given
temperature, provided we know the values of the exchange couplings
$J_{1a},J_{1b},J_2,J_c$ and the single-ion anisotropy $J_s$.

When the temperature $T$ approaches zero, we obtain
$\coth(\frac{\beta\omega_{\bm k}}{2})\to 1$. The equation
(\ref{eq06}) is reduced to
\begin{equation}\label{eq08}
\Phi|_{T\to0}=\frac{1}{2N}\sum_{\bm k}\left[\frac{A_{\bm
k}}{\sqrt{A_{\bm k}^2-B_{\bm k}^2}}-1\right].
\end{equation}
The zero-temperature average spin $\langle S^z\rangle_0$ can be
obtained by self-consistently solving the equations
(\ref{eq05})(\ref{eq07})(\ref{eq08}).

When the temperature $T$ approaches the magnetic transition
temperature $T_N$, the average spin $\langle S^z\rangle$ as well as
the spectrum $\omega_{\bm k}$ will approach zero. Expanding
$\coth(\frac{\beta\omega_{\bm k}}{2})$ in the equation (\ref{eq06}),
we obtain
\begin{equation}\label{eq09}
\Phi|_{T\to T_N}\approx\frac{\Gamma}{\beta\langle
S^z\rangle}-\frac{1}{2},
\end{equation}
where $\Gamma=\frac{1}{N}\sum_{\bm k}\frac{A_{\bm k}}{A_{\bm
k}^2-B_{\bm k}^2}$. Inserting (\ref{eq09}) into (\ref{eq05}), and
expanding the terms in the denominator and the numerator as the
series of $\langle S^z\rangle$, we finally derive
\begin{equation}
\langle S^z\rangle\approx\sqrt{\frac{12(\Gamma k_B
T_N)^2}{S(2S-1)}\left(1-\frac{T}{T_N}\right)},
\end{equation}
where
\begin{equation}\label{eq10}
T_N=\frac{S(S+1)}{3k_B\Gamma}.
\end{equation}
On the other hand, inserting (\ref{eq09}) into (\ref{eq07}), and
using the equation (\ref{eq10}), we obtain the reduced expression
for $\Theta^{(z)}$ near the temperature $T_N$:
\begin{equation}
\Theta^{(z)}|_{T\to T_N}\approx\frac{2S-1}{3S}.
\end{equation}
\section{Numerical results and discussions}\label{sec3}
So far there is no consensus on the magnitudes of  the exchange
couplings $J_{1a},J_{1b},J_2$ and $J_c$, because of the unclear
microscopic origin of the observed AF spin structure. Here we prefer
the estimations in ref\cite{Liu}, which gave $J_{1b}=50\pm10$ meV,
$J_{1a}=49\pm 10$ meV, $J_2=26\pm5$ meV and $J_c=0.020\pm0.015$ meV
by using the experimental transition temperature $T_N=138$K of pure
LaFeAsO. The main purpose of this paper is to investigate the
magnitude and the effects of the single-ion anisotropy $J_s$ in
LaFeAsO. Through out our numerical calculation, we take $J_{1b}=50$
meV, $J_{1a}=49$ meV, $J_2=26$ meV and the spin $S=1$. The result
$J_{1b}\sim50$meV are obtained from the first-principle
calculating\cite{Ma,Yin}. In the present systems of units, the
Boltzmann constant is taken as $k_B=0.086$ meV/K.

\begin{figure}
  \begin{center}
  \includegraphics[scale=0.9]{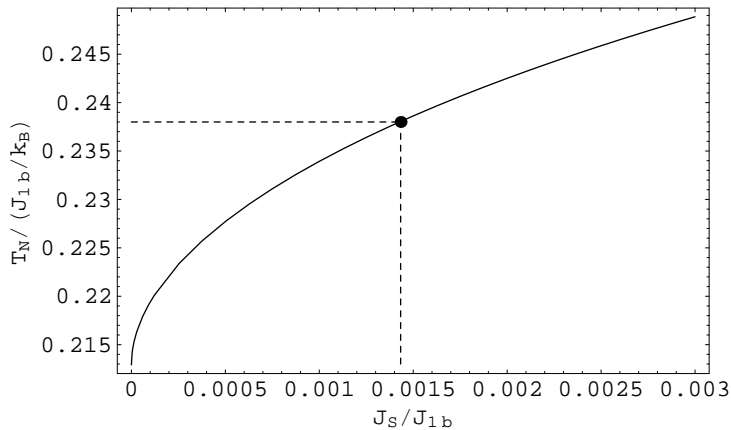}
  \end{center}
  \caption{\label{fig2}The transition temperature $T_N$
  as a function of the single-ion anisotropy $J_s$ for $J_{1a}=0.98, J_2=0.52$
  and $J_c=0.0004$ in the unit of $J_{1b}=50$ meV. The black point correspond to
  the experimental temperature $T_N=0.238$ in the unit of $J_{1b}/k_B$.}
\end{figure}

\begin{figure}
  \begin{center}
  \includegraphics[scale=0.8]{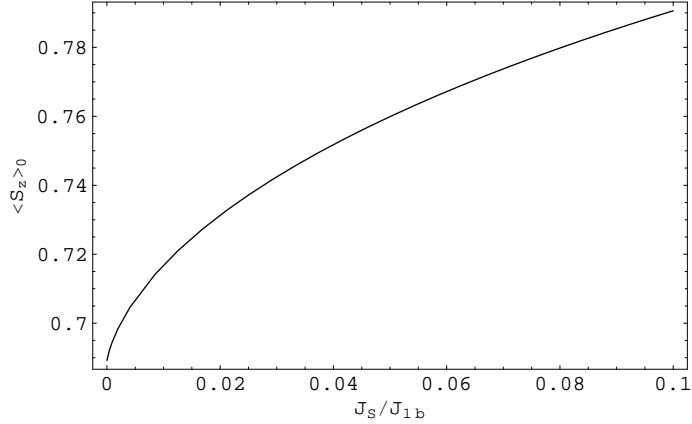}
  \end{center}
  \caption{\label{fig3}The zero-temperature average spin as
  a function of the single-ion anisotropy $J_s$ for $J_{1a}=0.98$, $J_2=0.52$ and $J_c=0.0004$
  in the unit of $J_{1b}=50$ meV.}
\end{figure}

Figure \ref{fig2} shows the effect of the single-ion anisotropy
$J_s$ on the transition temperature $T_N$. We can see that $T_N$
increases as $J_s$ increases. This means that the single-ion
anisotropy term is in favor of the AF spin structure. It can be
understood from the expression of the single-ion anisotropy term in
the Hamiltonian (\ref{eq01}). Increasing the magnitude of $J_s$ will
make the spins incline to align along the $z$ axis, and give a lower
total energy, which make the system more stable. To one's surprise,
the magnitude of $J_s$ corresponding to the experimental transition
temperature $T_N$ is about $0.00143J_{1b}$, which is much bigger
than the magnitude of the exchange coupling $J_c=0.0004J_{1b}$
estimated in ref\cite{Liu}. Furthermore, we see from figure
\ref{fig3} that the variation range of $\langle S^z\rangle_0$ with
$J_s$ varying from 0 to 0.1 is almost the same as the one produced
by $J_c$ in ref\cite{Liu}. All these results imply that the
magnitude of $J_s$ is probably not much small compared with $J_c$.
So the estimation of $J_c$ maybe need to be adjusted if the
single-ion anisotropy term is considered. As to the estimations of
the other exchange couplings $J_{1a}, J_{1b}$ and $J_2$, we think
there are still reasonable.

\begin{figure}
  \begin{center}
  \includegraphics[scale=0.9]{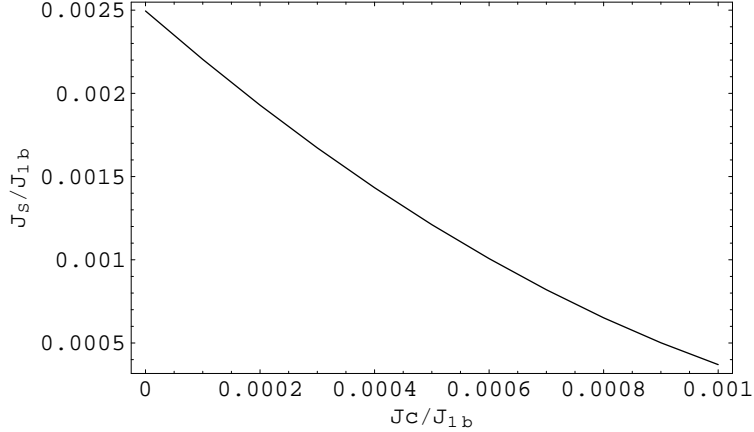}
  \end{center}
  \caption{\label{fig4}The competing relation of $J_c$ and $J_s$. The curve is depicted
  by using the equation (\ref{eq10}) for $T_N=0.238(J_{1b}/k_B)$, $J_{1a}=0.98J_{1b}$
   and $J_2=0.52J_{1b}.$}
\end{figure}
Figure \ref{fig4} shows the competing relation of $J_c$ and $J_s$
when the transition temperature is fixed at the experimental value.
The increase of $J_c$ is accompanied by the decrease of $J_s$, and
vice vera. From figure \ref{fig4}, we can see that the ranges of
their corresponding variations are at the same magnitude, which
implies they probably have the same status in the viewpoint of
theoretical study. As to revealing the actual magnitudes of the two
parameters $J_c$ and $J_s$, we think it is not enough to use only
the experimental transition temperature $T_N$.

\begin{figure}
  \begin{center}
  \setlength{\unitlength}{0.5cm}
  \includegraphics[scale=1.0]{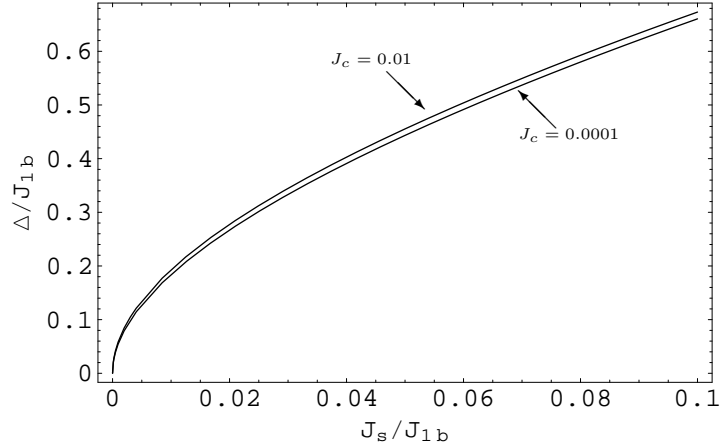}
  \put(-10.5,10.2){\tiny$J_c=0.01$}
  \put(-9,10){\vector(1,-1){1}}
  \put(-5.5,8.2){\tiny$J_c=0.0001$}
  \put(-4.5,8.5){\vector(-1,1){1}}
  \end{center}
  \caption{\label{fig5}The spin spectrum gaps at low temperature $T\to0$
  as functions of the single-ion anisotropy $J_s$
   for $J_{1a}=0.98J_{1b}$, $J_2=0.52J_{1b}$ and
   $J_c=0.0001J_{1b}$(the below curve), $J_c=0.01J_{1b}$(the above curve).}
\end{figure}
Figure \ref{fig5} shows the effect of the single-ion anisotropy
$J_s$ on the spectrum gap at low temperature. The gap vanishes as
$J_s$ vanishes, and increases as $J_s$ increases. We find the effect
of $J_c$ on the gap is very trivial. The curves for different values
of $J_c$ between $(0.0001,0.01)$ are almost the same, while the
single-ion anisotropy affects the gap apparently. Considering the
gap can be obtained from inelastic neutron-scattering
experiment\cite{Zhao}, we suggest that the magnitude of the
single-ion anisotropy be estimated from the future experimental
results of the spectrum gap. For example, if $J_c=0.0004J_{1b}$, we
obtain $J_s=0.00143J_{1b}$ from the experimental transition
temperature $T_N=138$K. Then calculating the spectrum gap with
$J_s=0.00143J_{1b}$, we obtain the magnitude of the gap
$\Delta\approx3.4$ meV, which can be compared with the future
experimental result.
\section{Conclusion}\label{sec4}
We use Green's function method to study the Heisenberg model
(\ref{eq01}) of LaFeAsO with the striped AF spin structure as shown
in figure \ref{fig1}. The main purpose of this paper is to
investigate the magnitude and the effects of the single-ion
anisotropy $J_s$. We derive the self-consistent equations for the
average spin, and obtained the analytical expressions for the spin
spectrum gap $\Delta$, and the magnetic transition temperature
$T_N$. We find that the transition temperature $T_N$, the
zero-temperature average spin $\langle S^z\rangle_0$ and the spin
spectrum gap $\Delta$ are all increasing functions of the single-ion
anisotropy $J_s$. From our numerical results by using $T_N=138$K and
the previous estimations of $J_{1a},J_{1b},J_2$ and $J_c$ in
ref\cite{Liu}, we find that the magnitude of $J_s$ is probably not
much small compared with $J_c$. Because the single-ion anisotropy is
essential for the existence of the spin spectrum gap, we suggest
using the experimental result of the spin spectrum gap to fix the
magnitude of the single-ion anisotropy $J_s$ in the future.


\section*{References}
\begin{thebibliography}{10}
\bibitem{Kami}Kamihara Y, Watanabe T, Hirano M and Hosono H 2008 {\it J.
Am. Chem. Soc.} {\bf 130} 3296
\bibitem{Yild}Yildirim T 2008 {\it Phys. Rev. Lett.} {\bf 101} 057010
\bibitem{Cao}Cao C, Hirschfeld P J and Cheng H P 2008 {\it Phys. Rev.} B
{\bf 77} 220506(R)
\bibitem{Ma}Ma F and Lu Z Y 2008 {\it Phys. Rev.} B {\bf 78} 033111
\bibitem{Dong}Dong J \etal 2008 {\it Europhys. Lett.} {\bf 83} 27006
\bibitem{Cruz}de la Cruz C \etal 2008 {\it Nature} {\bf 453} 899
\bibitem{Chen}Chen Y \etal 2008 {\it Phys. Rev.} B {\bf 78} 064515
\bibitem{Huang}Huang Q \etal 2008 {\it Phys. Rev. Lett.} {\bf 101} 257003
\bibitem{Zhao2}Zhao J \etal 2008 {\it Phys. Rev.} B {\bf 78} 140504(R)
\bibitem{Gold}Goldman A I \etal 2008 {\it Phys. Rev.} B {\bf 78} 106506(R)
\bibitem{McGu}McGuire M A \etal 2008 {\it Phys. Rev.} B {\bf 78} 094517; arXiv:0804.0796
\bibitem{Si}Si Q and Abrahams E 2008 {\it Phys. Rev. Lett.} {\bf 101} 076401
\bibitem{Fang}Fang C \etal 2008 {\it Phys. Rev.} B {\bf 77} 224509
\bibitem{Zhao}Zhao J \etal 2008 {\it Phys. Rev. Lett.} {\bf 101} 167203
\bibitem{Liu}Liu G B and Liu B G 2009 {\it J. Phys.:Condens. Matter}
{\bf 21} 195701
\bibitem{Yao}Yao D X and Carlson E W 2008 {\it Phys. Rev.} B {\bf 72} 052507
\bibitem{Klau}Klauss H H \etal 2008 {\it Phys. Rev. Lett.} {\bf 101} 077005
\bibitem{Tyab}Tyablikov S V 1959 {\it Ukr. Mat. Zh.} {\bf 11} 289; S. V. Tyablikov 1967
{\it Methods in the Quantum Theory of Magnetism} (New York: Plenum
Press)
\bibitem{Ander}Anderson F B and Callen H B 1964 {\it Phys. Rev.} {\bf 136} A1068
\bibitem{Frob}Fr\"{o}brich P, Jensen P J and Kuntz P J 2000 {\it Eur.
Phys. J.}B {\bf 13} 477
\bibitem{Hene}Henelius P, Fr\"{o}brich P, Kuntz P J \etal 2002
{\it Phys. Rev.} B {\bf 66} 094407
\bibitem{Call}Callen H B 1963 {\it Phys. Rev.} {\bf 130} 890
\bibitem{Yin}Yin Z P \etal 2008 {\it Phys. Rev. Lett.} {\bf 101} 047001
\end{thebibliography}
\end{document}